%% file: paper.tex
\documentclass[11pt,dvips]{article}
\usepackage{epsfig}
\usepackage{wrapfig}
\begin{document}
\begin{center}


{\LARGE \bf Determination of the EAS Attenuation Length from Data 
of the ANI Experiment} 
\end{center}

\begin {center}
{
\bf A.A. Chilingarian, G.V. Gharagyozyan, S.S. Ghazaryan, 
G.G. Hovsepyan, \\ E.A. Mamidjanyan,
L.G. Melkumyan, S.H. Sokhoyan\footnote{corresponding 
author: e-mail: serg@crdlx5.yerphi.am}\\
}
{\it{Yerevan Physics Institute, Cosmic Ray Division, Armenia.}}
\end{center}

\vspace{1ex}
\begin{center}
\begin{minipage}[c]{5.5in}
Using the EAS size spectra measured with the MAKET ANI 
array on Mt. Aragats, Armenia ($3200m$ a.s.l.- $700 g\cdot cm^{-2}$) 
in the range of $N_e = 10^5 - 10^7$ for different 
angles-of-incidence, the EAS attenuation length has been 
determined applying different analysis methods. 
The analysis is based on a data sample of $2.5 \cdot 10^6$ events 
collected in the period of June, 97 - April, 99. 
The results are compared with results deduced from data of the 
EAS TOP and KASCADE experiments.
\end{minipage}
\end{center}
\vspace{1ex}

\section {Introduction}
The intensity of Extensive Air Showers (EAS) with fixed shower 
sizes $N_e$ is assumed to decrease exponentially with increasing  
atmospheric depth of the observation level. This is considered to 
be due to the absorption of the particles of the EAS 
cascade following an exponential law 
\begin{equation}\label{cascade} 
N_e(X)=N_e(X_0)exp{\left(-\frac{X-X_0}{\Lambda}\right)},  
\,\,\,\, {\rm with}\,\, X \geq X_0 .
\end{equation}
$X_0$ is a definite initial atmospheric depth after the maximum 
of the longitudinal development where the number of (charged) 
particles is $N_e(X_0)$ and further decreasing exponentially, 
$N_e(X)$ is the number of particles of the EAS at the slant 
depth $X[g \cdot  cm^{-2}]$. \\
The quantity $\Lambda$ controls the attenuation of particles of the individual cascade 
\cite{hayakawa} {\it (size attenuation 
length)}. It is related to the inelastic cross sections (to the 
mean free path length $\lambda_A$) of the interaction of the 
primary cosmic ray particles with air nuclei. The attenuation of 
the flux intensity of Extensive Air Showers is characterized by 
a related quantity $\lambda_N$ {\it (intensity attenuation length, absorption)}, 
which can be directly measured by cosmic rays detector arrays. 
Thus measurements of the attenuation of the EAS intensity in the 
atmosphere are considered to be an interesting source of 
information about hadronic interactions, especially if extended  
to the ultrahigh energy region expected 
from the forthcoming LHC and TESLA accelerators. In addition due 
to the sensitivity of the cross sections to the mass of the 
primary, alterations of the attenuation length with the energy 
may be indicative for the variations of the mass composition. 
Measured results imply tests of the energy dependence of the 
extrapolated cross sections used for Monte Carlo simulations. \\
The investigations of the present paper are based on an EAS  
sample measured 1997-1999 with the MAKET ANI array
\cite{avak,hovsep98} on Mt. Aragats station (Armenia) and 
registered for different angles-of-incidence in the zenith angle 
interval $\Theta = 0 - 45^\circ$. The data basis of the analysis 
can be enlarged by published data from KASCADE $(1046\, g \cdot cm^{-2})$ 
\cite{kascade} and EAS TOP $(810\, g \cdot cm^{-2})$ \cite{top1} 
experiments. Spectra measured by EAS TOP are given in Ref. 
\cite{top2}. Data and zenith angle dependence for KASCADE results 
are obtained by scanning the spectrum plots communicated by 
the KASCADE collaboration \cite{ralf1}. \\
We apply different procedures to deduce the attenuation. 
First we consider the degradation of the EAS flux with fixed 
shower size $N_e$ with increasing zenith angle i.e. increasing 
atmospheric thickness of the shower development (characterized 
by the intensity attenuation length ($\lambda_N$) \cite{khris}). 
Differently the technique of the constant intensity cut (CIC) 
\cite{nagano} considers the intensity spectrum of EAS events 
and relates equal intensities observed at different atmospheric 
depths. 

There is the tacit assumption that the shower size 
reflects the energy of the primary. The procedure can be 
refined by using the knee position in the $N_e$ spectrum as a 
bench mark for a well defined energy, so far we may associate 
the knee phenomenon to a feature of the primary energy 
spectrum of cosmic radiation.

\section{Experimental spectra} 
The experimental basis of the present investigations are 
measurements of shower size spectra in the knee region and their 
zenith-angle dependence performed with the MAKET ANI array of the 
Mt. Aragats Cosmic Ray Station (3200 m a.s.l.) in Armenia. 
Details of the measurements and the experimental procedures 
taking into account the detector response are given elsewhere 
\cite{gagik,hovsep99}. For a detailed description of the knee 
region the traditional approximation with two different spectral 
indices below and above the knee, defining the knee position as 
intersection of two lines in a logarithmic presentation, 
appears to be insufficient. Hence a more sophisticated method 
has been applied with parameterization of the slope of the spectra 
(see Ref.\cite{serg1}). \\ 
Tab.1 compiles the characteristics of the size spectra measured 
with the MAKET ANI installation, the changes of the slopes in the 
knee region ($\Delta N_{e_k}$), expressed by different spectral 
indices below ($\gamma_ 1$) and  above ($\gamma_ 2$) the knee 
position $N_{e_k}$ for the zenith-angle range of $\Theta = 0 -
45^\circ$. For the display and the analysis of the zenith-angle 
dependence, the size spectra are determined in 5 angular bins 
of equal $\Delta\sec\Theta$ widths. The accuracy of the zenith 
angle determination is estimated to be about $1.5^\circ$ 
\cite{gagik}. A correction due to barometric pressure changes, 
which lead to small fluctuations of the atmospheric absorption, 
has not been made. Figure \ref{makasdif} displays the spectra 
of mean values of each atmospheric depth bin and compares with 
the results from EAS-TOP \cite{top2} and KASCADE \cite{ralf1} 
experiments. \\
\input gsum

\begin{figure}
\begin{minipage}[t]{0.48\linewidth}
\epsfig{file=spcdifall.epsi,width=8.cm,height=8.cm} 
\vspace{-.5cm}
\caption{\label{makasdif} \it{Differential size spectra for 
different zenith angles ranges observed with MAKET ANI array,  
compared with spectra reported by the KASCADE \cite{ralf1} and 
the EAS TOP \cite{top2} collaborations.}}
\end{minipage} 
\hspace*{0.4cm}
\begin{minipage}[t]{0.48\linewidth}
\epsfig{file=lincas8.epsi,width=8.cm,height=8.cm}
\vspace{-.3cm}
\caption{\label{lincas8} \it{$N_e$ cascade in the 
observed range of the atmospheric slant depth.}}
\vspace{.5cm}
\end{minipage}
\end{figure}
Following fixed intensities of the experimental spectra 
(see sect.3.2) the average $N_e$ cascade development can be 
immediately reconstructed as shown in Figure \ref{lincas8}. 
Note that the results in the range of the slant depth observed 
with the ANI array deviate from the exponential decrease 
(eq.\ref{cascade}). That is an interesting feature which can be 
revealed more clearly when combining spectra accurately measured 
on different altitudes. In the present paper we base the 
formulation of the procedures estimating the attenuation on the 
exponential decrease (eq.\ref{cascade}). It is our interest to 
explore, if this assumption applied to the ANI and KASCADE data 
lead to consistent results.
 
\section{Procedures for inference of the attenuation length \\
from size spectra}
We consider the differential and integral size spectra 
$I(N_e,X)$ and $I(>N_e,X)$, respectively. 
In addition to the basic assumption of exponential attenuation  
of $N_e$ (eq.\ref{cascade}) a power-law dependence of the size 
spectrum
\begin{equation}
I(N_e,X)\propto{N_e}^{-\gamma},
\end{equation}
with the spectral index $\gamma$ is adopted.

\subsection{\boldmath Attenuation of the intensity of fixed $N_e$: 
absorption length}
For different fixed values of shower size $N_e$, on different 
depths in the atmosphere or/and different zenith angles of 
incidence, from measured spectra (see vertical dotted lines on 
Figure \ref{spintgcic}) we obtain several values of corresponding 
intensities from the equivalent depths from 700 till 1280
$g\cdot cm^{-2}$. 
Fitting the depth dependence of the intensities by the straight 
line (in logarithmic scale) according to equation:
\begin{equation}
I(N_e,X)= I(N_e,X_0)exp\left(-\frac{X-X_{0}}{\lambda_N}\right) 
\label{eq-abs}  
\end{equation}
we obtain the estimate of the absorption length $\lambda_N$. 
The absorption length can be estimated both by integral and
differential spectra.
\begin{figure}
\begin{minipage}[b]{0.48\linewidth}
\vspace{.2cm}
\epsfig{file=spintgmakas.epsi,width=8.cm,height=8.cm}
\vspace{-.2cm}
\caption{\label{spintgcic} \it{Integral size spectra for different 
zenith angles ranges observed with MAKET ANI array, compared with 
spectra reported by the KASCADE \cite{ralf1}: illustration of
the procedures for absorption and attenuation length estimates.}}
\end{minipage}
\hspace*{0.4cm}
\begin{minipage}[b]{0.48\linewidth}
\epsfig{file=xkndifintg.epsi,width=8.cm,height=8.cm}
\vspace{-.2cm}
\caption{\label{xkn} \it{The variation of the knee position 
with the atmospheric depth.}}
\vspace{1.5cm}
\end{minipage}
\end{figure}
\subsection{Constant intensity cut}
The basic idea of this procedure is to compare the average size of 
showers which have the same rate (showers per $m^2 \cdot s \cdot sr$)
in the different bins of the zenith angle of shower incidence and 
different slant depth, respectively \cite{nagano}. \\
Considering two different depths in atmosphere 
$X_{1}$,$X_{2} > X_{0}$ the expressions of differential
intensities $I(N_e,X)$ has the form
\begin{equation}\label {eq-cic}
N_e(X_1)^{-\gamma}exp\left[-\left(\gamma-1\right)\frac{X_1-X_{0}}{\Lambda}
\right]=N_e(X_2)^{-\gamma}exp\left[-\left(\gamma-1\right)\frac{X_2-X_{0}}{\Lambda}\right]
\end{equation}
With simple transformations we obtain:
\begin{equation}\label {cicdif}
\Lambda_{diff}(I)=\frac{\gamma-1}{\gamma}\frac{X_2-X_1}
{ln\left(\frac{N_e(X_1)}{N_e(X_2)}\right)}
\end{equation}
The attenuation lengths, obtained by integral spectra do not depend 
explicitly on spectral index:
\begin{equation}\label {cicint}
\Lambda _{int}(I)=\frac{X_2-X_1}{ln\left(\frac{N_e(X_1)}{N_e(X_2)}\right)}
\end{equation} 
Practically the estimate of the attenuation length is obtained by 
fitting the $N_e$ dependence on the depth in atmosphere by the 
straight line according to the equation (\ref {cascade}). 
The sequence of $N_e$ values is obtained according to the fixed 
values of the flux intensity, selected from the interpolation of 
the differential or integral size spectra. \\
For each $N_e$ value, the slope index $\gamma$  used in equation 
\ref{cicdif}, is obtained by averaging over all used slant depths.
Selecting equal intensities ($\approx$ primary energies) 
corresponding to different shower sizes $N_e$ and different 
depths the value of $\Lambda_{diff} (I)$ is estimated. 
Intensity values from $10^{-9}$ to $5.\cdot 10^{-6}$
were used for CIC method.

\subsection{Attenuation of the size of the knee}
A special variant of the constant intensity cut is to follow the 
decrease of the shower size at a constant primary energy in the 
size spectrum. Assuming that the knee phenomenon reflects a 
feature of the primary flux, the variation shower size at the 
knee with the zenith angle provides the possibility to extract 
the  attenuation length. \\
Considering the assigned knee position of the data from various
experiments, differences within 30\% are noticed for all X-bins. \\
The knee positions obtained by the differential and integral 
spectra are a bit shifted to the smaller $N_e$ values 
(see Figure \ref{xkn}). The shift is approximately uniform over 
all investigated depths interval, therefore the estimates of 
the attenuation length by the differential and integral 
size spectra are very close to each other.

\subsection{The relation between the absorption and 
attenuation length}
We consider the quantity $I(N_e,X)dN_e$ - the number of EAS at
the depth $X$ which comprise $N_e$ to $N_e+dN_e$ particles: 
\begin{equation}
I(N_e,X)dN_e\sim N_e^{-\gamma}exp\left[-\left(\gamma-1\right)
\frac{X-X_{0}}{\Lambda}\right]dN_e  
\end{equation}
With eq.\ref{eq-abs} we obtain:
\begin{equation}
\Lambda_{diff}(N_e) =  (\gamma(N_e)-1)\lambda_N,
\end{equation}
where, $\gamma(N_e)$ is the differential size spectra index 
(here we indicate the $N_e$ dependence of the slope index explicitly).  
For the integral spectra: 
\begin{equation}\label {attfromabs}
\Lambda_{int}(N_e) =  \gamma(N_e)\lambda_N,
\end{equation}
where, $\gamma(N_e)$ is integral size spectra index. \\  
For the evaluation of the inelastic cross section and for 
comparison of the three methods described above we propose to 
use the calculated values of the attenuation length $\Lambda$
(instead of using absorption length $\lambda_N$).
The attenuation of the number of particles in the individual 
cascade is more directly connected with the characteristics of the 
strong interaction and is independent from the parameters of the
cosmic ray flux incident on the atmosphere. In turn the absorption 
length, i.e. the attenuation of the CR flux intensity, reflects also 
characteristics of the primary flux and is
dependent on the change of the slope of the spectra. 

\subsection{Estimate of the inelastic cross section} 
The inelastic cross sections, of the primary 
nuclei with atmosphere nuclei is related by \cite{nagano}: 
\begin{equation}
\sigma^{inel}_{A-air}(mbarn) = 
\frac{2.41\cdot 10^4}{\lambda_A(g\cdot cm^{-2})}, 
\end{equation}
where A denotes the primary nuclei. The quantity $\lambda_A$ is 
the {\it interaction length} of the A-nucleus in the atmosphere 
(note: in some publications the interaction length is 
denoted by $\lambda_N$, where N is primary nuclei, in contrast in this paper N is  
reserved for the shower size). 
The interaction length 
$\lambda_A$ is related with the absorption length $\Lambda_A$ by  
\begin{equation}
\lambda_A = K(E) \cdot \Lambda_A
\end{equation}
The K(E) coefficient reflects 
peculiarities of the strong interaction model used for simulation.
The value of the parameter K has to be determined by 
simulations of the EAS development in the atmosphere. 
Such studies require the development of procedures for the  
selection of EAS initiated by primaries of a definite type 
(see for example in \cite{beam1,beam2}). 

\section{Application to the data}
\label{disc.sec}
The mean values of the attenuation lengths obtained by various  
methods from data of the ANI and KASCADE installations, 
as well as for the joint ANI \& KASCADE data by
the differential ($\Lambda_{diff}$) and integral spectra 
($\Lambda_{int}$) are compiled in the Tables 
\ref{gtab1},\ref{gtab3},\ref{gtab2}.
\vspace{0.15cm}
\input gtab 
The alternative estimates of the attenuation length 
reflect the inherent uncertainties of the methods 
and the statistical errors, as well as the fluctuations of cascade 
development in the atmosphere, the 
\begin{figure}
\begin{minipage}[b]{0.48\linewidth}
\epsfig{file=lambintg3fit.epsi,width=8.cm,height=8.cm}
\vspace{-.4cm}
\caption{\label{lamintgcas3} \it{Attenuation Length dependence 
on Spectra Intensity (Primary Energy).}}
\end{minipage}
\hspace*{0.4cm}
\begin{minipage}[b]{0.48\linewidth}
\vspace{-.5cm}
\epsfig{file=lamgtintg.epsi,width=8.cm,height=8.cm}
\vspace{-.4cm}
\caption{\label{lamdif9} \it{Attenuation Length dependence on 
the Shower Size $N_e$.}}
\end{minipage}
\end{figure}
\begin{wrapfigure}[24]{r}{9.cm}
\vspace{-0.1cm}
\epsfig{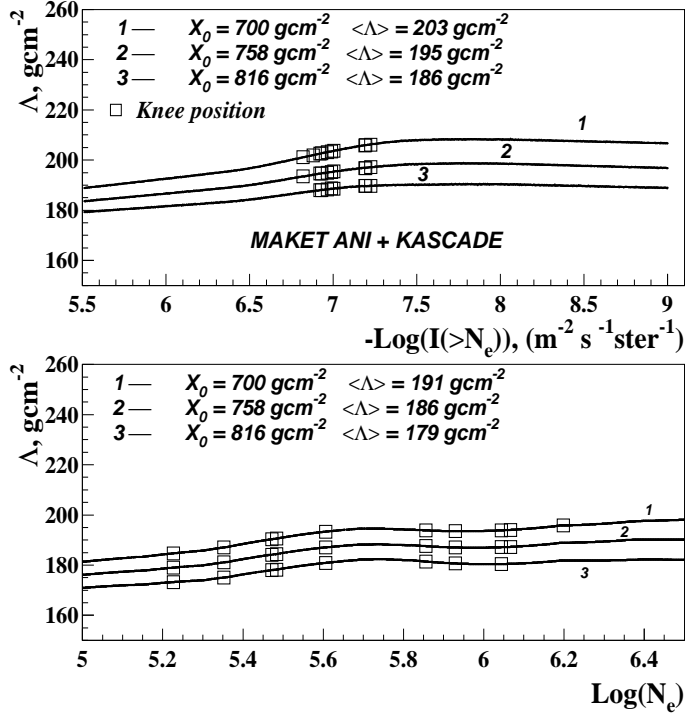}
\vspace{-0.4cm}
\caption{\label{makas2} {\it Attenuation length
obtained by joint analysis of the MAKET ANI and KASCADE data.}}
\end{wrapfigure}
energy dependence of 
the inelastic cross section and possible changes in mass composition.
As obvious in Figure \ref{lincas8}, the values
corresponding to the minimal equivalent depths of  used MAKET ANI data, deviate 
significantly from the exponential decrease.
The observations reflects the flattening of the cascade curve 
just after the shower maximum in the altitude $500-600\,g\cdot cm^{-2}$. 
Due to these features the attenuation lengths calculated by 
MAKET ANI data appear to be 
significantly larger than those derived for the KASCADE 
data (Tables \ref{gtab1}, \ref{gtab3}). \\
Therefore, for the combined analysis of the KASCADE and ANI data 
we omitted the first and the second zenith
angle bins of MAKET ANI and calculate
the attenuation lengths by the remaining 9 (minimal equivalent depth 
$758\,g\cdot cm^{-2}$) and 8 (minimal equivalent depth $816\,g\cdot cm^{-2}$) 
angular bins. The dependences of estimated values of attenuation 
length on the shower size and flux intensity for different 
amount of the angular bins used, are displayed in Figures 
\ref{lamintgcas3} (note, that higher intensities on the X 
axes correspond to the lower primary energies) and \ref{lamdif9}.    
The attenuation length estimates obtained from the differential 
and integral spectra agree fairly well.
The results of both CIC and recalculation from absorption 
length agree within the error bars. 
The results obtained by the "attenuation of knee position" are 
larger for MAKET ANI and KASCADE.
As pointed out by S. Ostapchenko \cite{ostap} it is the consequence of 
the large EAS fluctuations with the tendency to
shift the knee position to the lower energies 
(and correspondingly to higher fluxes) in a way to "slow down"
the cascade curve attenuation. \\
Well below the shower development maximum starting from 
$816 g\cdot cm^{-2}$ KASCADE and MAKET ANI
data could be fitted with one decay parameter
(see Figure \ref {makas2}). There is a concentration of the 
knee positions on the curve showing the dependence of 
the attenuation of the flux
intensity ($\approx$ primary energy). 
In turn, the curve displaying the dependence of the
attenuation length on the shower size demonstrates a
rather large dispersion of the "knee positions". These
observations in size
and energy scales may be interpreted as an indication of the
astrophysical nature of the knee phenomenon.

\section{Conclusion}\label{conc.sec}
Experimental studies of EAS characteristic like the depth of 
the shower maximum $X_{max}$, the elongation rate 
$dX_{max}/dlog_{10}E$ and the attenuation length $\Lambda$ are of 
particular importance, since they map rather directly basic 
features of the hadronic interaction. Strictly, however, the 
interpretation of these quantities in terms of hadronic cross 
sections cannot bypass the necessity of detailed calculations 
of the shower development. Nevertheless these type of EAS 
quantities, if compared with Monte Carlo simulation results, 
provide stringent tests of the interaction model ingredients of 
the simulations. \\
The recent results of various experimental installations are 
sufficiently accurate to enable relevant studies of this kind, 
and combining the data from arrays situated on different 
altitudes (like MAKET ANI and KASCADE) allows a large span in 
the atmospheric slant depth for reconstructing the development 
of the charge particle size. In fact such studies, if using a 
sufficiently large data sample, could  be continued in a more 
detailed manner by separating the muon component and taking 
into account the deviations from the exponential shape of 
the cascade decline. The penetrating muon component  
contributes with smaller attenuation to the development of 
the considered charged particle component, but hardly with 
an exponential degrading (according to eq.\ref{cascade}). 
Actually by use of methods in progress to isolate different 
primary groups ("pure nuclear beams") of the size spectra 
\cite{beam2,gagik99}, these kind of interaction studies would get 
of extreme interest. 

 \vspace*{3.5ex}
{\noindent \bf \Large Acknowledgment \\}
\vspace*{-.5ex}

{\it \noindent This publication is based on experimental 
results of the ANI collaboration. The MAKET ANI detector installation has been 
set up as collaborative project of the Yerevan Physics Institute,
Armenia and the Lebedev Institute, Moscow. The continuous 
contributions and assistance of the Russian colleagues in operating 
the detector installation and in the data analyses are gratefully 
acknowledged. In particular, we thank S. Nikolski and V. Romakhin 
for their encouraging interest and useful discussions. \\
First perspectives of combined considerations of the KASCADE 
and ANI experimental data have been discussed in 1998 during the 
ANI-98 workshop in the cosmic ray observatory station Nor-Amberd 
of Mt.Aragats (Armenia). The MAKET ANI group would like to thank 
the German colleagues for stimulating discussions and encouragement, 
in particular H. Rebel for his numerous valuable comments and 
interesting suggestions to the topic of this paper. We acknowledge 
the useful discussions with K.-H. Kampert, H. Klages and 
R. Glasstetter. The suggestions of S. Ostapchenko are highly 
appreciated. \\  
The work has been partly supported by the research grant 
No.96-752 of  the Armenian Government and by the ISTC project A116.  
The assistance of the Maintenance Staff of the Aragats Cosmic Ray 
Observatory in operating the MAKET ANI installation is 
highly appreciated.}


\end {document}

%% file: gsum.tex
\begin{table}[h]
\def\arraystretch{1.2}
\centering
\begin{tabular}{ll}
 
\hline
$I(10^5<N_e<1.15\cdot10^6)$      & $(8.95\pm 0.18)\cdot10^{-11}(N_e/10^5)^{\gamma_1}$\\
$I(N_e>2.56\cdot10^6)$                & $(3.23\pm 0.40)\cdot10^{-13}(N_e/10^6)^{\gamma_2}$\\
$\gamma_1$                            &  $-2.54\pm 0.012 $\\
$\gamma_2$                            &  $-2.94\pm 0.042$ \\
$\Delta (N_{e_k})$                  &  ($1.15\pm 0.034)\cdot10^6$ \ - \ $(2.56\pm 0.063)\cdot10^6$ \\
$N_{e_k}$                                &  $(1.75\pm 0.05) \cdot 10^6$\\
$I(N_{e_k})$                            &  $(5.83\pm 0.14)\cdot 10^{-14}$\\
\hline
\end{tabular}
\caption{\it Flux $[m^{-2}s^{-1}sr^{-1}]$ and knee region parameters 
of the size spectra measured with the MAKET ANI array.} 
\label{gsum1}
\end{table}

%% file: gtab.tex
 
\begin{table}[h]
\centering
\begin{tabular}{|c|c|c|c|c|c|c|}
\hline
  Min.depth & \multicolumn{2}{c|}{MAKET ANI} & 
\multicolumn{2}{c|}{ANI+KASCADE}& \multicolumn{2}{c|}{KASCADE} \\
\hline
$X_0,g\cdot cm^{-2}$ & $\Lambda_{int}$ & $\Lambda_{dif}$ & $ \Lambda_{int}$ & 
$ \Lambda_{dif}$ & $\Lambda_{int}$ & $\Lambda_{dif}$ \\
\hline
$ 700$ & $248\pm27$ & $247\pm42$ & $203\pm10$ & $203\pm13$ & $ - $ & $ - $\\ 
\hline
$ 758$ & $236\pm32$ & $237\pm51$ & $195\pm8$ & $196\pm12$ & $ - $ & $ - $\\ 
\hline
$ 816$ & $211\pm43$ & $218\pm70$ & $186\pm9$ & $188\pm13$ & $ - $ & $ - $\\ 
\hline
$1020$ & $ - $ & $ - $ & $ - $ & $ - $ & $181\pm14$ & $182\pm23$\\ 
\hline
\end{tabular}
\caption{\it Attenuation lengths for the data from the 
MAKET ANI and KASCADE installations estimated
 by the CIC method from differential and integral size spectra}
 \label{gtab1}
\end{table} 
\begin{table}[h]
\vspace{-0.15cm}
\centering
\begin{tabular}{|c|c|c|c|c|c|c|}
\hline
  Min.depth & \multicolumn{2}{c|}{MAKET ANI} & 
\multicolumn{2}{c|}{ANI+KASCADE}& \multicolumn{2}{c|}{KASCADE} \\
 \hline
$X_0,g\cdot cm^{-2}$ & $\Lambda_{int}$ & $\Lambda_{dif}$ & $ \Lambda_{int}$ &
$\Lambda_{dif}$ & $\Lambda_{int}$ & $\Lambda_{dif}$ \cr
\hline
$ 700$ & $239\pm14$ & $240\pm15$ & $191\pm11$ & $193\pm13$ & $ - $ & $ - $\\ 
\hline
$ 758$ & $232\pm13$ & $228\pm19$ & $186\pm10$ & $184\pm17$ & $ - $ & $ - $\\ 
\hline
$ 816$ & $213\pm14$ & $219\pm27$ & $179\pm11$ &  $181\pm24$ & $ - $ & $ - $\\ 
\hline
$1020$ & $ - $ & $ - $ & $ - $ & $ - $ & $181\pm7$ & $183\pm11$\\ 
\hline
\end{tabular}
\caption{\it Attenuation lengths for the data from the 
MAKET ANI and KASCADE installations estimated
 by the recalculation from the absorption length for differential and 
 integral size spectra}
\label{gtab3}
\end{table}
\begin{table}[h]
\vspace{-0.15cm}
\centering
\begin{tabular}{|c|c|c|c|c|c|c|} 
\hline
  Min.depth & \multicolumn{2}{c|}{MAKET ANI} & 
\multicolumn{2}{c|}{ANI+KASCADE}& \multicolumn{2}{c|}{KASCADE} \\
\hline
$X_0,g\cdot cm^{-2}$ & $\Lambda_{int}$ & $\Lambda_{dif}$ & $ \Lambda_{int}$ &
$\Lambda_{dif}$ & $\Lambda_{int}$ & $\Lambda_{dif}$ \cr
\hline
$ 700$ & $302\pm71$ & $295\pm83$ & $241\pm17$ & $237\pm15$ & $ - $ & $ - $\\ 
\hline
$ 758$ & $272\pm51$ & $263\pm42$ & $242\pm20$ & $221\pm17$ & $ - $ & $ - $\\ 
\hline
$ 816$ & $ - $ & $ - $ & $225\pm21$ &  $225\pm19$ & $ - $ & $ - $\\ 
\hline
$1020$ & $ - $ & $ - $ & $ - $ & $ - $ & $232\pm26$ & $222\pm28$\\ 
\hline
\end{tabular}
\caption{\it Attenuation lengths for the data from the MAKET ANI and
KASCADE installations, estimated
 by the "attenuation of knee position" method from differential and integral size spectra}
\label{gtab2}
\end{table}

%% file: paper.bbl
\vspace*{-0.3cm}   

\begin{thebibliography}{99}
\renewcommand{\baselinestretch}{0.1}
\parskip0.ex
%
\bibitem{hayakawa}
S. Hayakawa, {\it Cosmic Ray Physics}, Interscience Monographs and 
Texts in Physics and Astronomy, V. 22, Wiley-Interscience, 1969
\bibitem{avak}
V.V. Avakyan et al., Jadernaya Fiz. 56 (1993) 182 
\bibitem{hovsep98}
G.G. Hovsepyan for the ANI collaboration., Proc. of the Workshop 
ANI 98, eds. A.A. Chilingarian, H.Rebel, M. Roth, M.Z. Zazyan, 
FZKA 6215, Forschungszentrum Karlsruhe 1998, 45
\bibitem{kascade}
H.O. Klages et al. - KASCADE collaboration,  
Nucl. Phys. B (Proc. Suppl.) 52B (1997) 92  
\bibitem{top1}
M. Aglietta et al., Nucl. Instrum. and Meth. A336 (1993) 310 
\bibitem{top2}
M. Aglietta et al., Astropart. Phys. 10 (1999) 1 
\bibitem{ralf1}
R. Glasstetter et al. - KASCADE collaboration, Proc. 16{th} 
ECRS (Alcala, 1998), 564 
\bibitem{khris}
G.B. Khristiansen, G. Kulikov, J. Fomin,  
{\it Cosmic Rays of Superhigh Energies}, Verlag Thiemig, M\"unchen,
1979
\bibitem{nagano}
M. Nagano et al., Journ. Phys. G: Nucl. Phys. 10 (1984) L235; \\
Gaisser T.K,  {\it Cosmic Rays and Particle Physics}, 
Cambridge Univ. Press, 1992  
\bibitem{gagik}
G.V. Gharagyozyan for the ANI collaboration, Proc. of the Workshop ANI 98, eds.
A.A. Chilingarian, H.Rebel, M. Roth, M.Z. Zazyan, FZKA 6215,
Forschungszentrum Karlsruhe 1998, 51  
\bibitem{hovsep99}
S.V. Blokhin, V.A. Romakhin, G.G. Hovsepyan, 
these proceedings
\bibitem{serg1}
S.H. Sokhoyan S.H. et al. - ANI collaboration, 
Proc. of the Workshop ANI 98, eds.
A.A. Chilingarian, H.Rebel, M. Roth, M.Z. Zazyan, FZKA 6215,
Forschungszentrum Karlsruhe 1998, 55
\bibitem{beam1}
A.A. Chilingarian, H.Z.Zazyan, Yad. Fiz. 54 (1991) 128
\bibitem{beam2}
A. Vardanyan et al., these proceedings 
\bibitem{ostap} 
S. Ostapchenko , private communication, 1999
\bibitem{gagik99}
G.V. Gharagyozyan et al., these proceedings   

\end{thebibliography}
